\documentclass[a4paper,11pt,nofootinbib]{article}
\usepackage{pos}

\usepackage{caption}

\usepackage{comment}
\usepackage[caption = false]{subfig}

\usepackage{mathrsfs}
\usepackage{amsmath}
\usepackage{mathtools}
\usepackage{physics}
\usepackage{slashed}
\newcommand{\nc}{N_\mathrm{c}}
\newcommand{\aem}{\alpha_\mathrm{em}}
\newcommand{\gev}{\mathrm{GeV}}
\newcommand{\jpsi}{$\mathrm{J}/\psi$}
\newcommand{\jpsim}{\mathrm{J}/\psi}
\newcommand{\bt}{\mathbf{b}}
\newcommand{\rt}{\mathbf{r}}
\newcommand{\Bt}{\mathbf{b}}
\newcommand{\xpom}{x_\mathbb{P}}
\newcommand{\pt}{\mathbf{p}}
\newcommand{\xt}{\mathbf{x}}
\newcommand{\yt}{\mathbf{y}}
\newcommand{\Deltat}{\mathbf{\Delta}}
\newcommand{\xbj}{x_\mathrm{Bj}}
\newcommand{\ft}{F_{2}}
\newcommand{\fl}{F_{\mathrm{L}}}
\newcommand{\nf}{n_\mathrm{f}}
\newcommand{\seq}{\sum_{q}^{\nf} e_q^2}
\newcommand{\lqcd}{\Lambda_\text{QCD}}
\newcommand{\cf}{C_\mathrm{F}}
\newcommand{\TR}{T_\mathrm{R}}
\newcommand{\seqav}{ \bar{e}_q^2}
\newcommand{\cgf }{ C}
\newcommand{\csf }{ C}
\newcommand{\cflglo}{ {C}_{\fl g}^{(1)}}
\newcommand{\cflqlo }{ {C}_{\fl q}^{(1)}}
\newcommand{\cflslo }{ {C}_{\fl \Sigma}^{(1)}}
\newcommand{\cflgnlo}{ {C}_{\fl g}^{(2)}}
\newcommand{\cflnsnlo }{ {C}_{\fl \rm NS}^{(2)}}
\newcommand{\cflpsnlo }{ {C}_{\fl \rm PS}^{(2)}}
\newcommand{\cftgnlo}{ {C}_{\ft g}^{(1)}}
\newcommand{\cftqnlo }{ {C}_{\ft q}^{(1)}}
\newcommand{\cfkgnlo}{ {C}_{\fk g}^{(1)}}
\newcommand{\cfkqnlo }{ {C}_{\fk q}^{(1)}}
\newcommand{\wft}{\widetilde{F}_2}
\newcommand{\cflwft}{ {C}_{\fl \wft}}
\newcommand{\wftp}{\widetilde{F'}_2}
\newcommand{\wfl}{\widetilde{F}_{\mathrm{L}}}
\newcommand{\wflp}{\widetilde{F'}_{\mathrm{L}}}
\newcommand{\wflpp}{\widetilde{F''}_{\mathrm{L}}}
\newcommand{\ftw }{F_2^{\rm W^-}}
\newcommand{\wftw}{\widetilde{F}_2^{\rm W^-}}
\newcommand{\fk }{F_3}
\newcommand{\fkw }{F_3^{\rm W^-}}
\newcommand{\ftcw }{F_{2\rm c}^{\rm W^-}}
\newcommand{\wftcw}{\widetilde{F}_{2\rm c}^{\rm W^-}}
\newcommand{\xq }{ (x, Q^2)}
\newcommand{\eu }{ e_u^2}
\newcommand{\ed }{ e_d^2}
\newcommand{\es }{ e_s^2}

\newcommand{\as}{\alpha_\mathrm{s}}
\newcommand{\As}{\frac{\as}{2\pi}}
\newcommand{\der}{\mathrm{d}}

\title{Evolution of structure functions at NLO without PDFs}

\author[a,b]{Tuomas Lappi}
\author[a,b]{Heikki Mäntysaari}
\author[a,b]{Hannu Paukkunen}
\author[a,b]{Mirja Tevio}

\affiliation[a]{Department of Physics, University of Jyväskyla,  P.O. Box 35, 40014 University of Jyväskyla, Finland}

\affiliation[b]{Helsinki Institute of Physics, P.O. Box 64, 00014 University of Helsinki, Finland}

\emailAdd{tuomas.v.v.lappi@jyu.fi}
\emailAdd{heikki.mantysaari@jyu.fi}
\emailAdd{hannu.t.paukkunen@jyu.fi}
\emailAdd{mirja.h.tevio@jyu.fi}

\abstract{We formulate the Dokshitzer-Gribov-Lipatov-Altarelli-Parisi (DGLAP) evolution of the Deep Inelastic Scattering (DIS) structure functions $F_2$ and $F_{\rm L}$ at next to leading order in $\alpha_s$ (NLO) directly in terms of the structure functions rather than parton distributions (PDFs). We call this the physical basis approach. 
In practice, we first express the NLO quark singlet and gluon PDFs in terms of the structure functions $F_2$ and $F_{\rm L}$ in  momentum space. Employing these expressions in  the DGLAP evolution, we arrive at the evolution equations for $F_2$ and $F_{\rm L}$ in the physical basis. We demonstrate how one is free from defining a factorization scale and scheme when using the physical basis evolution equations. We also discuss the process of applying the NLO physical basis to global analysis of LHC cross sections.   }

\FullConference{31st International Workshop on Deep Inelastic Scattering (DIS2024)\\
 8–12 April 2024\\
Grenoble, France\\}

\begin{document}
\maketitle

\setlength{\abovedisplayskip}{3pt}
\setlength{\belowdisplayskip}{3pt}

\section{Introduction}
\addtocounter{page}{0}

The future Electron-Ion-Collider (EIC)~\cite{AbdulKhalek:2021gbh} will measure Deep Inelastic Scattering (DIS) cross sections, increasing the importance for the accurate DIS structure function predictions. Parton distribution functions (PDFs) are widely used un-observable quantities for expressing QCD processes. However, they are dependent on the arbitrary factorization scale and scheme, which adds a theoretical uncertainty to the predictions of QCD observables. 

In an alternative approach one formulates the DGLAP evolution of DIS structure functions directly in the so called physical basis. The physical basis consists of linearly independent DIS structure functions -- instead of PDFs -- and is therefore free from the factorization scheme and scale dependence. On the other hand, it is also more straightforward to directly parametrize physical observables when fitting data. The idea of a physical basis was already discussed about forty years ago in Ref.~~\cite{Furmanski:1981cw}, and more recently for example in Refs.~\cite{Catani:1996sc,Blumlein:2000wh,Hentschinski:2013zaa,Harland-Lang:2018bxd,Blumlein:2021lmf,vanNeerven:1999ca,RuizArriola:1998er}. The novelty in this work is that the final results are expressed in momentum space, instead of Mellin space, and the final physical basis is a full three flavour basis at next to leading order in $\as$ (NLO). This work is continuation for our previous work on the LO physical basis, which was published earlier this year~\cite{Lappi:2023lmi}.

\section{Constructing a physical basis}
\subsection{Two observable physical basis}

In order to understand the method of constructing the physical basis, it is easiest to first consider a basis consisting of only two observables. We choose to construct the two observable physical basis with the structure functions $\ft$ and $\fl$, which are related to PDFs by convolutions with coefficient functions $C_{F_{2,L} f_j}$
\begin{equation}
\label{eq:F2L}
    F_{2,L}\xq = \sum_{j}C_{F_{2,L} f_j}(Q^2, \mu^2)\otimes f_j(\mu^2), 
\end{equation}
where the PDFs $f_j(\mu^2)$ are the quark singlet over light flavours 
$ \Sigma(x, \mu^2) = \sum_{q}^{\nf}\left[q(x, \mu^2)+\overline{q}(x, \mu^2)\right]$, with $\nf=3$, and the gluon PDF $g(x,\mu^2)$.
The first step towards the physical basis is to invert the linear mapping from the PDF basis to the basis of structure functions, i.e. expressing the quark singlet and the gluon in terms of the structure functions as
\begin{equation}  
\label{eq: PB PDFs}
f_j(\mu^2) = \sum_{i}C_{F_i f_j}^{-1} (Q^2, \mu^2)\otimes F_i(Q^2)+\mathcal{O}(\as^2).
\end{equation}
Here the expressions have been truncated at the order $\as^2$, which will be discussed in more detail in the next section.

\begin{center}
\begin{figure}[b!]
\centering
    \subfloat[$\ft$ CT14 NLO]{%
    \includegraphics[width=0.5\textwidth]{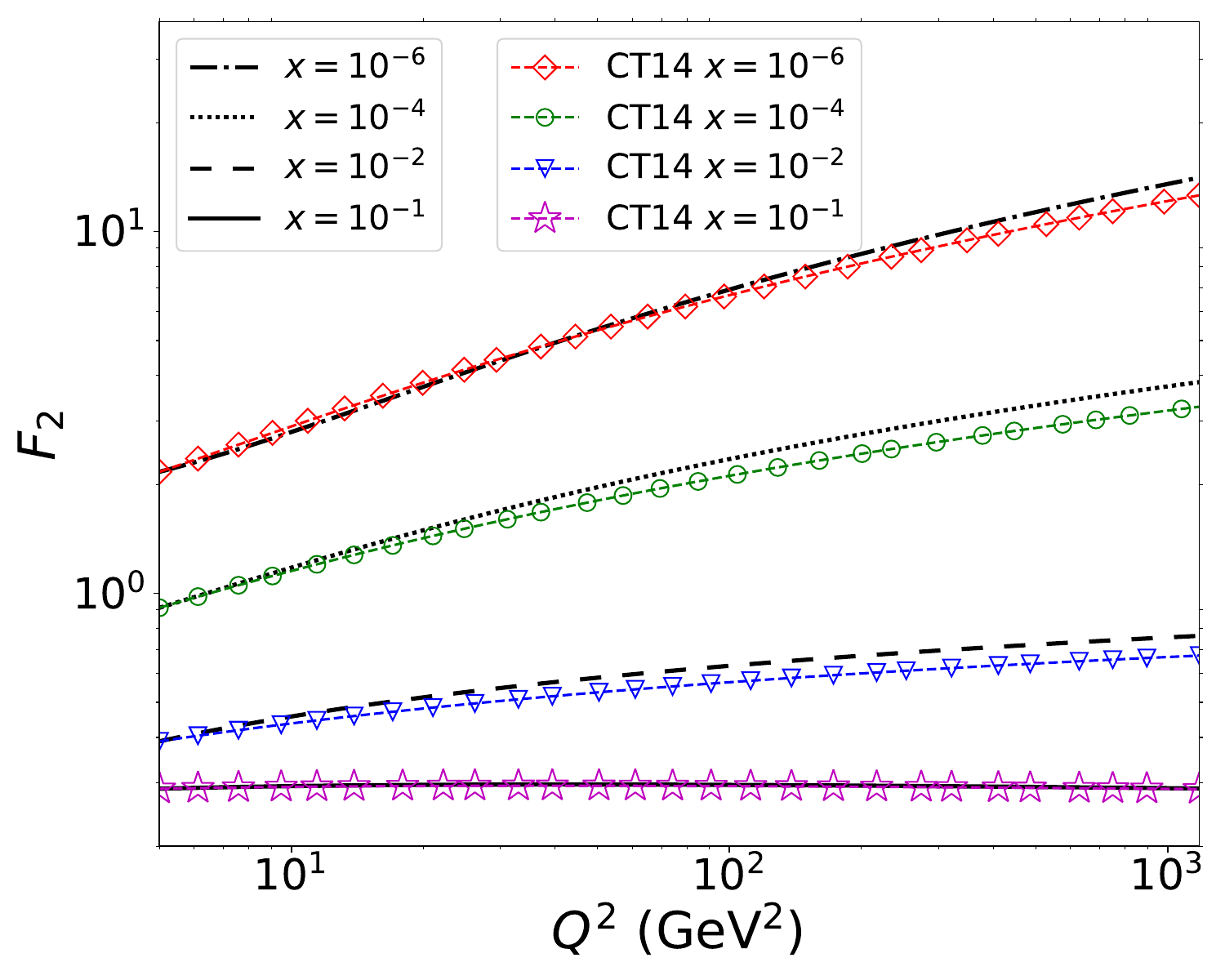}
    }
    \subfloat[$\fl$ CT14 NLO]{%
        \includegraphics[width=0.5\textwidth]{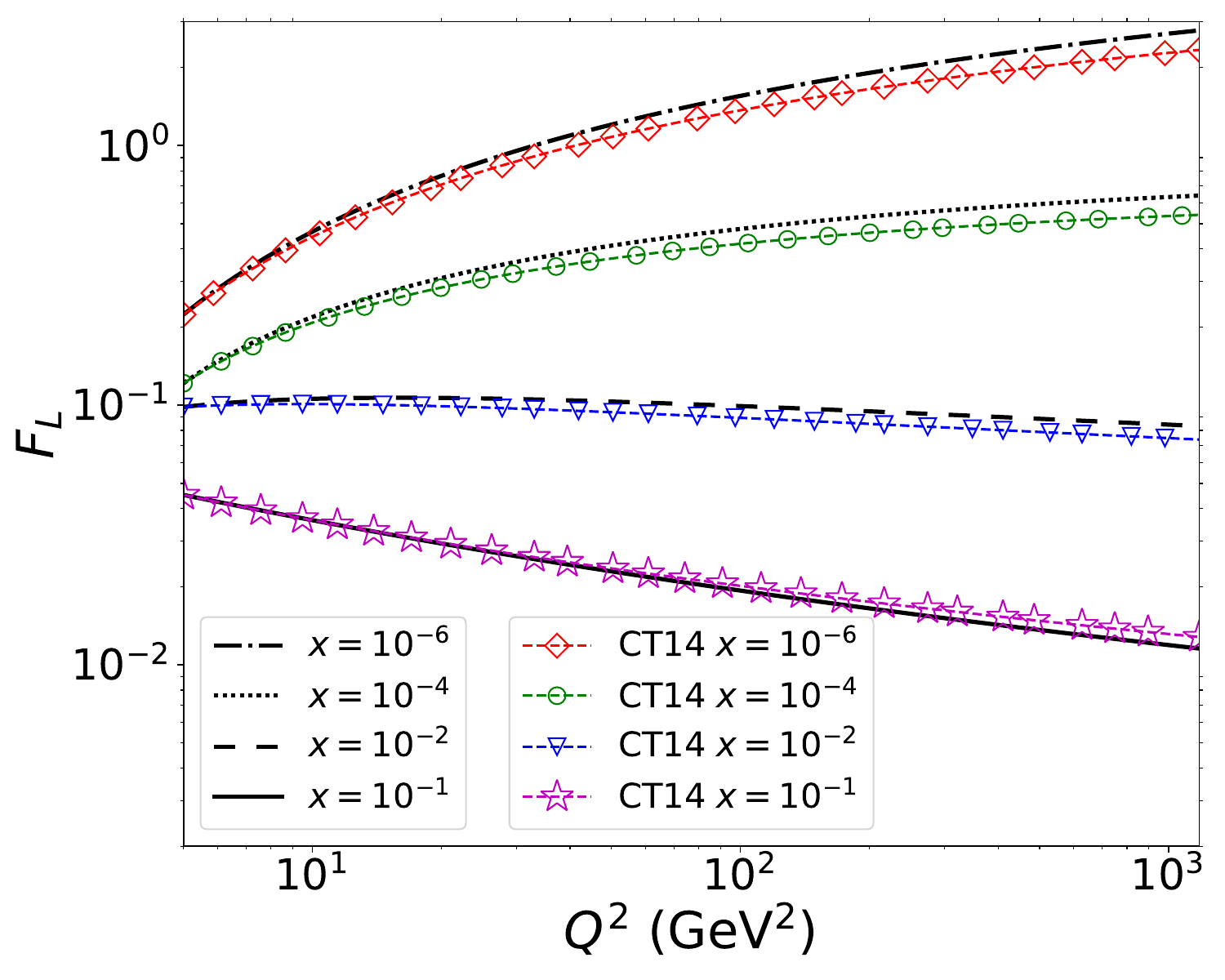}
    }
    \caption{Comparison of the structure functions $\ft$ and $\fl$ computed from DGLAP evolved PDFs (colourful markers) to the structure functions computed via DGLAP evolution in physical basis (black lines). Here the initial values for the physical basis evolution are calculated using PDFs. For PDFs we used the LHAPDF library~\cite{Buckley_2015} and the CT14lo\_NF3 PDF set.  }
\label{fig:1}
\end{figure}

\end{center}

 When constructing the DGLAP evolution in the physical basis, we start from the compact notation of the conventional form of DGLAP evolution
 \begin{equation}
\begin{aligned}
\label{eq: DGLAP}
     \frac{\dd F_i\xq}{\dd \log(Q^2)} &  = \sum_{j}\frac{\dd C_{F_i f_j}(Q^2, \mu^2)}{\dd \log(Q^2)}\otimes f_j(\mu^2) ,
\end{aligned}
\end{equation} 
where the DGLAP splitting functions are hidden inside the $Q^2$ derivatives of the coefficient functions. We then move to evolution in the physical basis just by inserting the expressions for PDFs in the physical basis in Eq.~\eqref{eq: PB PDFs}
\begin{equation}
\begin{aligned}
\label{eq: PB DGLAP}
     \frac{\dd F_i\xq}{\dd \log(Q^2)} &  = \sum_{j}\frac{\dd C_{F_i f_j}(Q^2, \mu^2)}{\dd \log(Q^2)}\otimes  \sum_{k}C_{F_k f_j}^{-1} (Q^2, \mu^2)\otimes F_k(Q^2) +\mathcal{O}(\as^3) \\ & \equiv \sum_k \mathcal{P}_{ik} \otimes F_k(Q^2)+\mathcal{O}(\as^3),
\end{aligned}
\end{equation}
where $i=2,L$ and $f_j= \Sigma, g$. Here we have truncated the expression at $\mathcal{O}(\as^3)$. From Eq.~\eqref{eq: PB DGLAP} one can see that the factorization scheme and scale dependence has to cancel within the evolution kernels $\mathcal{P}_{ik}$, and therefore we do not have to fix the value for the factorization scale $\mu$. The cancellation has to happen between the coefficient functions and the splitting functions from which the evolution kernels are composed of.   
We have implemented the two dimensional physical basis numerically, and the results are shown in Fig.~\ref{fig:1}. As expected at NLO, one sees difference in between the DGLAP evolved values in the physical basis and the PDF based values. The differences comes from the uncertainty caused by the scheme and scale dependence in the PDFs, and also from the perturbative truncation in the inversion from the structure functions to the PDFs in Eq.~\eqref{eq: PB PDFs}.

\subsection{Inverting the linear mapping perturbatively}

The inversion of the linear mapping in Eq.~\eqref{eq: PB PDFs} can be done exactly at the leading order (LO) in $\as$, by first inverting PDFs in the Mellin space, and then identifying the momentum space solution. In the next to leading order (NLO) in $\as$ the coefficient functions in Eq.~\eqref{eq:F2L} are not as simple as in LO, thus an exact inversion becomes challenging. However, we can invert the linear mapping perturbatively. 

A straightforward demonstration of the perturbative inverting is to consider a case with only the gluon PDF in the following expression of the structure function $\fl$
\begin{equation}
\begin{aligned}
\label{eq: inverting FL}
    \wfl = \cflglo\otimes g
    + \frac{\as(Q^2)}{2\pi} C_{\fl g}^{(2)}\otimes g,
\end{aligned}
\end{equation}
where 
    $ \wfl\xq \equiv (2\pi / \as(Q^2)) \fl(x, Q^2)/x$.
If one now defines a differential operator 
\begin{equation}
\label{eq:P}
    \hat{P}(x)  \equiv \frac{1}{8\TR\nf\seqav}\left[x^2\frac{\dd^2}{\dd{x}^2}-2x\frac{\dd}{\dd{x}} +2\right],
\end{equation}
and uses it to operate on the first term of right hand side of Eq.~\eqref{eq: inverting FL}, only the gluon PDF remains     
\begin{equation}
\begin{aligned}
\label{eq:g Phat 1}
   g(x) = \hat{P}(x)\left[\cflglo\otimes g\right].
\end{aligned}
\end{equation}
Now one can express the term $\cflglo\otimes g$ as 
   $\wfl - \frac{\as(Q^2)}{2\pi} C_{\fl g}^{(2)}\otimes g $ and insert that into Eq.~\eqref{eq:g Phat 1}, which leads to an expression 
    \begin{equation}
    \begin{aligned}
    \label{eq:g Phat 2}
       g(x) &=  \hat{P}(x)\left[\wfl(x) - \frac{\as(Q^2)}{2\pi} C_{\fl g}^{(2)}\otimes g \right],
    \end{aligned}
    \end{equation}
where one can replace the gluon on the right hand side by inserting  $g (x)= \hat{P}(x)\wfl(x)+\mathcal{O}\left(\as(Q^2)\right)$
\begin{equation}
\begin{aligned}
\label{eq:g Phat 2}
   g(x) &=   \hat{P}(x)\wfl(x) - \frac{\as(Q^2)}{2\pi} \hat{P}(x)\Big[  C_{\fl g}^{(2)}\otimes \hat{P}\wfl \Big]+ \mathcal{O}\left(\as^2(Q^2)\right),
\end{aligned}
\end{equation}
and truncating the solution at $\as^2$.

The method of pertubatively inverting the gluon PDF, shown above, can be extended to a system with all the quark flavours. One just needs to entail the same degrees of freedom in the physical basis as in the PDF basis. The perturbative expansion can also be continued to higher orders in $\as$. However in order to be consistent with the perturbative order of the physical basis, the truncation order should match the order of the selected structure functions. The perturbative inversion prevents calculating an exactly conserved momentum sum rule from the inverted PDFs.
    
\subsection{Extending to a six observable physical basis}

A simplified example of a physical basis with two observables was discussed above. As already mentioned, the same steps can be applied to establish a more complete physical basis. In our a work in progress we are constructing a six dimensional physical basis which covers the quark flavours $u$, $\bar{u}$, $d$, $\bar{d}$, and $s=\bar{s}$. By including also the gluon PDF in our set or partons, we have in total six degrees of freedom meaning that in order to obtain the physical basis we need to choose six linearly independent structure functions. From the neutral current DIS we choose structure functions $\ft$ and $\fl$ corresponding to the virtual photon exchange, and $\fk$ corresponding to the $Z$-boson exchange. Then from charged current DIS we choose structure functions $\ftw$, $\fkw$, and $\ftcw$ corresponding to the $W^-$ boson exchange. Here we do not consider the quark mixing.

\section{Cross sections in a physical basis}

Since the physical basis approach is based on replacing PDFs, at least in principle, one can express all the PDF dependent cross sections in a physical basis. Here we consider an example cross section; a Higgs production by gluon fusion, defined as 
\begin{equation}
    \label{eq:HiggsproductionPDFbasis}
    \sigma(p+p\longrightarrow {\rm H}+X) = \int\dd x_1\dd x_2  g(x_1,\mu)g(x_2,\mu) \hat \sigma_{gg \rightarrow H + X}\left(x_1, x_2,\frac{m_H^2}{\mu^2}\right),
\end{equation}
where $m_H$ is the Higgs mass, $\hat \sigma_{gg \rightarrow H + X}$ is the parton level cross section, and $g(x_1,\mu)$ and $g(x_2,\mu)$ are the gluon PDFs. Expressing the Higgs production cross section in terms of a physical basis is simple; one just plugs in the physical basis expression for the gluon PDF \\
$g(x,\mu^2) = \sum_j C_{jg}^{-1}(Q^2,\mu^2) \otimes F_j(Q^2) $, where $F_j = \ft, \text{ } \fl,\text{ }\fk,\text{ } \ftw,\text{ } \fkw, \text{ and } \ftcw$ in the six observable basis. The Higgs production cross section in terms of physical basis is then expressed as
\begin{equation}
\begin{aligned}
   & \sigma(p+p\longrightarrow H+X) =  \\& \int\dd x_1\dd x_2 \hat \sigma_{gg \rightarrow H + X}(x_1, x_2,\frac{m_H^2}{\mu^2})
    \left[\sum_j C_{jg}^{-1}(Q^2,\mu^2) \otimes F_j(Q^2)\right]_{x_1} 
    \left[\sum_k C_{kg}^{-1}(Q^2,\mu^2) \otimes F_k(Q^2)\right]_{x_2},
\end{aligned}
\end{equation}
where the subscripts $x_1$ and $x_2$ refer to the Bjorken-$x$ values in the convolutions. It was noticed in Ref.~\cite{Harland-Lang:2018bxd} that when cross sections are structured as above, the explicit $\mu$ dependence cancels, and in the terms containing logarithms only rations of physical scales $\log(Q^2/m_H^2)$ remain. 
\section{Summary}

We have constructed a two dimensional physical basis at NLO, for which we have made a numerical implementation. We have discussed on how the physical basis approach is free from the factorization scale and scheme dependence. The extension for the full three-flavour physical basis, with six observables, has been established formally; however, the numerical implementation is still in progress. 
We have demonstrated how the physical basis can be applied to other processes, such as Higgs production by fusion of two gluons. 

The future work will study LHC cross sections in terms of physical basis. The aim is also to implement heavy quark flavours in our approach, which will increase the number of the structure functions needed to span the physical basis.

\section*{Acknowledgements}
This work was supported by the Academy of Finland, the Centre of Excellence in Quark Matter (projects 346324 and 346326), projects 321840 (T.L, M.T), project 308301 (H.P., M.T), and projects 338263 and 346567 (H.M).
This work was also supported under the European Union’s Horizon 2020 research and innovation programme by the European Research Council (ERC, grant agreements No. ERC-2018-ADG-835105 YoctoLHC and ERC-2023-101123801 GlueSatLight) and by the STRONG-2020 project (grant agreement No. 824093). 
Views and opinions expressed are however those of the authors only and do not necessarily reflect those of the European Union or the European Research Council Executive Agency. Neither the European Union nor the granting authority can be held responsible for them.

\bibliographystyle{JHEP-2modlong.bst}
\bibliography{refs}

\providecommand{\href}[2]{#2}\begingroup\raggedright\begin{thebibliography}{10}

\bibitem{AbdulKhalek:2021gbh}
R.~Abdul~Khalek {\em et.~al.}, {\it {Science Requirements and Detector Concepts for the Electron-Ion Collider}: {EIC Yellow Report}},  \href{http://dx.doi.org/10.1016/j.nuclphysa.2022.122447}{{\em Nucl. Phys. A} {\bf 1026} (2022) 122447} [\href{http://arXiv.org/abs/2103.05419}{{\tt arXiv:2103.05419 [physics.ins-det]}}].

\bibitem{Furmanski:1981cw}
W.~Furmanski and R.~Petronzio, {\it {Lepton - Hadron Processes Beyond Leading Order in Quantum Chromodynamics}},  \href{http://dx.doi.org/10.1007/BF01578280}{{\em Z. Phys. C} {\bf 11} (1982) 293}.

\bibitem{Catani:1996sc}
S.~Catani, {\it {Physical anomalous dimensions at small x}},  \href{http://dx.doi.org/10.1007/s002880050512}{{\em Z. Phys. C} {\bf 75} (1997) 665} [\href{http://arXiv.org/abs/hep-ph/9609263}{{\tt arXiv:hep-ph/9609263}}].

\bibitem{Blumlein:2000wh}
J.~Blumlein, V.~Ravindran and W.~L. van Neerven, {\it {On the Drell-Levy-Yan relation to {$\mathcal{O}(\as^2)$}}},  \href{http://dx.doi.org/10.1016/S0550-3213(00)00422-3}{{\em Nucl. Phys. B} {\bf 586} (2000) 349} [\href{http://arXiv.org/abs/hep-ph/0004172}{{\tt arXiv:hep-ph/0004172}}].

\bibitem{Hentschinski:2013zaa}
M.~Hentschinski and M.~Stratmann, {\it {On the Practical Application of Physical Anomalous Dimensions}},  \href{http://arXiv.org/abs/1311.2825}{{\tt arXiv:1311.2825 [hep-ph]}}.

\bibitem{Harland-Lang:2018bxd}
L.~A. Harland-Lang and R.~S. Thorne, {\it {On the Consistent Use of Scale Variations in PDF Fits and Predictions}},  \href{http://dx.doi.org/10.1140/epjc/s10052-019-6731-6}{{\em Eur. Phys. J. C} {\bf 79} (2019)~no.~3 225} [\href{http://arXiv.org/abs/1811.08434}{{\tt arXiv:1811.08434 [hep-ph]}}].

\bibitem{Blumlein:2021lmf}
J.~Bl\"umlein and M.~Saragnese, {\it {The N$^3$LO scheme-invariant QCD evolution of the non-singlet structure functions $F_2^\mathrm{NS}(x,Q^2)$ and $g_1^\mathrm{NS}(x,Q^2)$}},  \href{http://dx.doi.org/10.1016/j.physletb.2021.136589}{{\em Phys. Lett. B} {\bf 820} (2021) 136589} [\href{http://arXiv.org/abs/2107.01293}{{\tt arXiv:2107.01293 [hep-ph]}}].

\bibitem{vanNeerven:1999ca}
W.~L. van Neerven and A.~Vogt, {\it {NNLO evolution of deep inelastic structure functions: The Nonsinglet case}},  \href{http://dx.doi.org/10.1016/S0550-3213(99)00668-9}{{\em Nucl. Phys. B} {\bf 568} (2000) 263} [\href{http://arXiv.org/abs/hep-ph/9907472}{{\tt arXiv:hep-ph/9907472}}].

\bibitem{RuizArriola:1998er}
E.~Ruiz~Arriola, {\it {NLO evolution for large scale distances, positivity constraints and the low-energy model of the nucleon}},  \href{http://dx.doi.org/10.1016/S0375-9474(98)00489-8}{{\em Nucl. Phys. A} {\bf 641} (1998) 461}.

\bibitem{Lappi:2023lmi}
T.~Lappi, H.~M\"antysaari, H.~Paukkunen and M.~Tevio, {\it {Evolution of structure functions in momentum space}},  \href{http://dx.doi.org/10.1140/epjc/s10052-023-12365-2}{{\em Eur. Phys. J. C} {\bf 84} (2024)~no.~1 84} [\href{http://arXiv.org/abs/2304.06998}{{\tt arXiv:2304.06998 [hep-ph]}}].

\bibitem{Buckley_2015}
A.~Buckley, J.~Ferrando, S.~Lloyd, K.~Nordström, B.~Page, M.~Rüfenacht, M.~Schönherr and G.~Watt, {\it {LHAPDF}6: parton density access in the {LHC} precision era},  \href{http://dx.doi.org/10.1140/epjc/s10052-015-3318-8}{{\em The European Physical Journal C} {\bf 75} (mar, 2015) }.

\end{thebibliography}\endgroup


\begin{thebibliography}{99}
\bibitem{...}
....

\end{thebibliography}
\end{document}